# The Rise of Jihadist Propaganda on Social Networks

Adam Badawy, Emilio Ferrara[*]

*University of Southern California, Information Sciences Institute*

[*]Corresponding author: emiliofe@usc.edu

Using a dataset of over 1.9 million messages posted on Twitter by about 25,000 ISIS members, we explore how ISIS makes use of social media to spread its propaganda and to recruit militants from the Arab world and across the globe. By distinguishing between violence-driven, theological, and sectarian content, we trace the connection between online rhetoric and key events on the ground. To the best of our knowledge, ours is one of the first studies to focus on Arabic content, while most literature focuses on English content. Our findings yield new important insights about how social media is used by radical militant groups to target the Arab-speaking world, and reveal important patterns in their propaganda efforts.

## The Rise of Jihadist Propaganda on Social Networks

## Introduction

Militant groups have long used traditional media and the Internet to disseminate information, spread their propaganda, and recruit potential militants (Cohen-Almagor, R. 2012). But no group to date has been as savvy in terms of its propaganda campaign and recruiting terrorists via the Internet, and specifically via social media platforms, as the Islamic State of Iraq and the Levant (ISIS) (Shane & Hubbard, 2014). ISIS used a panoply of platforms, such as: Facebook, Instagram, Tumbler, Ask.fm, and most prominently, Twitter to spread its message (Bodine-Baron et al 2016). The group uses a very successful strategy of having an "online battalion" or the so-called "the mujtahidun (industrious)", that is likely a small group of 500-2,000 active online members who post and retweet certain tweets to make these messages trending, increasing the group's exposure and outreach (Berger & Morgan, 2015).

Previous research worked on understanding ISIS's social media presence (Bazan, Saad, and Chamoun, 2015; Klausen, 2015; Rowe and Saif, 2016; Ferrara, 2017) and the process of online radicalization (Bermingham et al. 2009; Edwards and Gribbon, 2013; Torok, 2013) that potential recruits go through. In this paper, we aim at identifying the topics that ISIS focuses on and investigate how offline events affect ISIS's online rhetoric. We thus attempt to answer the following two questions: *(i)* What are ISIS members and sympathizers talking about on Twitter and how would we categorize their conversations? *(ii)* How are external events reflected in the ISIS Twittersphere? Using a dataset containing all the Twitter posts of about 25,000 ISIS members and sympathizers, totaling over 1.9 million tweets generated from January 2014 to June 2015, we analyzed the Arabic content of the tweets to answer these questions. The reason for exclusive focus on Arabic content is that the majority of ISIS members come from the Arab world; Arabic is the most used language among ISIS members and sympathizers (Magdy et al. 2016; Berger & Morgan, 2015); and prior studies rarely focused on the Arabic content due to the numerous technical challenges. Additionally, Arabic posts constitute over 92% of ISIS activity in our dataset. In this study, we adopt state-of-the-art machine learning tools paired with human annotations to overcome these computational limits.

Our study is not the first to be concerned with identifying the topics and rhetoric that radical groups come to focus on and adopt (Haverson et al. 2011, Lee and Leets, 2009, Leuprecht et al. 2010, Payne, 2009). Smith et al. (2008) found that violent groups tend to focus on in-group affiliation, having an impact and confidence in victory. Moreover, they describe themselves using morality, religion, and aggression rhetoric more than non-violent counterparts. Our study supports similar hypotheses: we will show that ISIS members talk mainly about topics related to violence and Islamic theology, with a strong sectarian tone in certain circumstances. Moreover, we will bring evidence that important offline events are strongly intertwined with online conversation, with certain topics dominating the conversation right before or after ISIS's activity spikes. For example, we will show that increases in the incitement of violence by ISIS supporters and sympathizers on Twitter correlates with certain types of violent offline events; we will notice a steep rise in the theological talk following ISIS self-proclaiming itself a caliphate; concluding, we will observe an increase in the sectarian tone when the Iranian and Iraqi governments engage with ISIS.

## Materials and Methods

### Data Collection

We used a dataset of accounts associated with ISIS members and sympathizers. The identification of these accounts has been manually performed and validated in two stages: first, a crowd-sourcing initiative called *Lucky Troll Club* leveraged hundreds of volunteer annotators with expertise in Arabic to identify suspicious accounts that could be tied to ISIS, and reported them to Twitter. The usernames and account IDs of these users were compiled in a publicly-available list. Twitter's anti-abuse team manually verified all suspension requests, and suspended all the accounts related to ISIS based on the violation of Twitter's Terms of Service policy against terrorist- or extremist-related activity (Ferrara et al. 2016). Of the *Lucky Troll Club* we retained only the accounts that have been actually suspended by Twitter via the double-verification mechanism. This yielded a little over 25,000 Twitter accounts associated with ISIS as an object of this study. These users were responsible for posting over 1.9 million tweets from January 2014 to June 2015. All these tweets were obtained through the Observatory on Social Media database (OSoMe) set up by Indiana University (Davis et al., 2016), which continuously collects the Twitter data stream from the Twitter API (a 10% random sample of the full Twitter data stream). This allowed us to obtain a large sample of the activity of these ISIS accounts, an important technical advantage since studies using small-sample data have shown well known biases (Morstatter et al., 2013).

### Data Analysis

The first step of data cleaning was to retain only tweets in Arabic. This process removed about 8% of the tweets in the original dataset, the remainder 92% being written in Arabic. We then processed the tweets by means of so-called tokenization, using the NLTK tokenizer (Nltk). Tokenization is the process of splitting a piece of text into individual words, namely word tokens (or tokens in short). The tokens undergo further processing: we used the ISRIStemmer to stem them (Taghva, 2005). Stemming is the process of reducing a word to its root (word stem). This will yield related words to map to the same stem. We removed all non-alphanumeric characters from the stemmed tokens, as well as all English alphabet, both in lower and upper case. Our goal is to identify the most commonly used stems in the ISIS tweets that have theoretical significance: In order to do this, we removed the stop words (using the stops words from (Alajmi, 2012) in **Table VII**) and then constructed a list of the top 100 stems used in the tweets under scrutiny.

Out of the 100 top stems, 66 stems do not convey a clear meaning of what the authors are saying, such as: (To go out/exit, خرج), (Land, ارض), and (Do/Does, هل), which leaves us with 34 stems that hold significant meanings. We decided to categorize these stems into four categories: violence, theological, sectarian, and names. The first three are the three main topics we found to be prevalent in the ISIS tweets. The last one, "Names", includes both names of individuals as well as the Islamic State. In **Tables I & II**, we provide the list of stems that make up each category, along with their frequency in the whole tweets corpus.

We decided to be as conservative as possible in terms of choosing which stems would be included in these categories. For example, "Violence" only includes stems that incite violent actions. "Theological", includes stems of words that are used frequently in Islamic theology or Fiqh. As for "Sectarian", we included only derogatory terms of other groups, such as: "رفض" for Rafidah (a derogatory term for Shia Muslims) (Stern & Berger 2015), "صفي" for Safawi (a derogatory term for Shia Muslims and it is the Arabized form of Safavid, the 16th century dynasty that established Iran as a Shia State), and "كفر" and "رتد" for infidel and apostasy respectively, which can be used as derogatory terms for Shias, Yazidis, Christians and even Muslims who do not adhere to ISIS's vision of Islam.

We used this dictionary-based approach to classify each tweet. We used this approach instead of a common alternative known as "bag-of-words"i because we believe that some individuals in our dataset wrote their messages in areas of conflict where the author might have written a short message instead of a longer one due to imminent danger. As such, classifying the tweets rather than the tokens enable us to avoid introducing a systematic bias due to exogenous events.

We built a tweet classifier that counts how many stems are present in each tweet that belongs to the four categories mentioned above. We applied a majority-rule to classify the tweets: for each tweet, the category with the most stems represented in that tweet would be chosen as the category of the tweet itself. If a tweet has equal and maximum number of stems from two categories, we classified it as "Other".ii Also, if a tweet has no stems from any of the categories, we classified it as "None" and excluded it from future analysis.

We isolated tweets classified as "Other," that contain at least one stem from the "Names" category. Since "Names" is the only category that does not have one defined theme, after removing the tokens classified under this category, we used the majority-rule again to classify the tweets as either "Violence", "Theological", or "Sectarian". We named these categories as "Names+Violence", "Names+Theological" and "Names+Sectarian" to identify three extra sets of tweets that concurrently identify named entities and one of the three other topics of discussion.

In **Table III**, we provide examples of the tweets classified under each category.

## Results

The classifier we built classified more than half of the tweets (56%) based upon the four categories mentioned in the data analysis section. In **Table IV**, we report the number of tweets classified under each category, the percentage of tweets classified under each category out of the whole dataset, and the percentage of the tweets classified under each category out of the tweets that were classified in one of the categories. As for the "Other" category (**Table V**), the amount of tweets that we were able to classify under one of the three subcategories "Names+Violence", "Names+Theological" and "Names+Sectarian", is 55%.

To represent the volume of the tweets, and how this changes over time, we plotted the ratio of each category to the whole number of the tweets in the dataset (**Figure I**). Moreover, for the "Names +

Category" dataset, we plotted the ratio of each category to the whole number of tweets in the "Names" category (**Figure II**).

From both figures, it appears evident that the time series fluctuate and spikes can be observed at different point in time for different thematic categories. Social media literature has reported on correlations between offline and online events in a variety of cases, including during events related to social issues and conflicts (Conover et al., 2013; Varol et al., 2014). We further investigated to see whether a relationship exists between certain offline events and the Twitter chatter among ISIS members and sympathizers. We created a list (**Table VI**) with the most important events related to the four categories of topics we outlined earlier and plotted each category separately with a customized subset of offline events, which are the most relevant to the selected category. **Figures III-VI** show the ratio of the tweets in the four categories over the total number of tweets by week and the corresponding offline events related to that category. We included similar plots for the "Names + Category" classes in **Figures VII-IX**.

## Discussion

More than half of the tweets in our dataset can be classified under the four categories we constructed: this suggests that a significant portion of chatter generated by ISIS members and sympathizers gravitates around the four issues outlined above. Two topics dominate the conversation, namely theological issues and violence. Looking at the share of these topics in the whole dataset and among those categorized suggests the importance of these two topics for ISIS: theological and violence related issues compose a little over 30% of all the tweets (and more than half of the ones classified). This fact warrants further discussion; thus, we try to draw some possible explanations.

Violence plays a major part in ISIS's brand and its appeal among ISIS followers. We suggest that ISIS transformed the goal of many Islamist groups. Since the foundation of al-Qaeda, Osama Bin Laden advocated focusing on the "head of the snake", as he called the USA. He is arguing that once one destroys the source of "evil" that dominates Muslim countries, its puppets (Arab and Muslim leaders) would consequently lose power, and Islamic lands would be freed. This message insisted on gradualism and portrayed the battle between Jihadists and the US and its supporters in the context of David vs. Goliath. On the other hand, ISIS, from the start, refused to portray itself as an underdog, focusing on its victories, atrocious violent acts against minorities (and particularly, Shias, whom they used as a target in order to gain support among certain sectors of the Sunni community), and its call for an Islamic state. This was an intentional posture, which is reflected profoundly in its message to: 1) excite and attract many young men and women to this "exciting" and "victorious" journey that brings pride to its participants, and 2) to escape the spiritual hegemony of al-Qaeda-central and Ayman al-Zawahiri's attempts to control the jihadi scene in Iraq and Syria. Therefore, we see a strong emphasis on violence-related topics in ISIS's Twittersphere.[iii] Our results confirm this possible explanation: we see a trend of violence-related rhetoric rising around some of the offline events we outlined. Four events stand out: the crucifixions in Raqqa (Abdelaziz, 2014), the invasion of Mosul and Tikrit (Saddem Hussein's hometown, a place of symbolic importance) by ISIS, the invasion of Sinjar (Yazidi town), and the taking over of the Shaer gas field. All these events, annotated in the timelines, align faithfully with conversation spikes online (**Figure III**). As for the "Names + Violence" tweets, we see the same pattern of relationship with the crucifixions (**Figure VII**). These results provide support for our prior expectations and theoretical interpretations.

For the theological tweets (**Figure IV**), we plotted the announcement of the caliphate and the sectarian events. Here, we expected a sharp rise in theological talk after the caliphate announcement, and suspected that ISIS, after inflicting violence upon minorities, would engage in theological defense and justifications for its actions. In line with what we hypothesized, we see a sharp increase in Twitter discussions among ISIS members and sympathizers after ISIS self-pronounced itself a caliphate, suggesting how important the perception of this event was among ISIS followers. For the "Names + Theological" tweets, we again see the same sharp rise in tweet volume after the announcement of the caliphate (**Figure VIII**). It is worth noting that ISIS's digital magazine *Dabiq* focused on legitimizing the caliphate since the announcement of its establishment (Winter, 2015). This should not be surprising, since the concept of the caliphate plays a crucial role in political Islam and arguably in Islamic theology or Fiqh. To many Islamists, Muslims need an "Islamic" state where they can live their lives under the guidance of the Sharia, or in their view free from exogenous corrupting influences. The need for an Islamic state, especially since the collapse of the Ottoman Empire, has been a motivating cause for many Islamic groups, including the groups that some might see as moderate, such as the Muslim Brotherhood.[iv]

For the "Sectarian" category plots (**Figure V**), we can see that the tweets under this category spiked in correlation with events that involves ISIS's sectarian enemies, most notably the Iranian and the Iraqi governments. For example, around mid-June of 2014, when Iran deployed forces to aid the Shia-dominated Iraqi government take back Tikrit and when the latter asked the US to conduct airstrikes against ISIS, we see very visible spikes in the ISIS Twittersphere that include derogatory terms regarding Shias and others in general. Another very clear spike in the usage of such derogatory terms happened the Iraqi government launched a massive campaign to get back Tikrit in March of 2015. For the "Names + Sectarian" class, we see the same pattern: events involving the Iranian or Iraqi government usually coincide with, or slightly precede, spikes in this kind of tweets (**Figure IX**).

Since the "Names" category includes names of various entities, we did not have a uniform expectation regarding the relationship of any specific offline event and the tweets classified under this category, except for the announcement of the caliphate. The "Names" category includes three stems related to the caliphate: a sharp rise in this category after the announcement of the caliphate by the leader of ISIS, Abu Bakr al-Baghdadi, appears as expected. (**Figure IV**).

## Conclusion

The chaos in the aftermath of the US invasion and withdrawal from Iraq and the popular uprising against Assad in Syria in 2011 created a vacuum that enabled groups like ISIS to form. But why did ISIS and not another group emerge as the most important and powerful organization in this context? A major reason behind that, is its spectacular ability to spread its violent and nihilistic message further and better than any of if its rivals, including established groups like, al-Qaeda. The most prominent way for ISIS to spread its propaganda is through online social platforms, most prominently Twitter. Thus, we attempted, using a dataset of millions of tweets posted by ISIS members and sympathizers during the one-year timeframe that witnessed ISIS's rise, to capture what these members are talking about, what message they wanted to convey and how events on the ground affect the Twittersphere. We concluded that violence, Islamic theology, and sectarianism play a crucial role in ISIS messaging. In some cases, ISIS emphasis on some topics of discussion slightly anticipated events on the ground: for example, the use of sectarian language online toward those entities perceived as adversaries was systematic prior to executions and attacks. In other cases, ISIS focused on certain topics as an aftermath of offline events: this was the case, for example, when ISIS inflicted violence upon minorities, and then engaged online in theological defense and justifications for its actions. Possibly, the most prominent event during this period, in terms of its importance and perceived meaning to ISIS sympathizers, was the announcement of the caliphate. This event was both preceded and followed by several shocks in the Twittersphere, with multiple spikes occurring across different categories of discussion shortly before and slightly after the event. In conclusion, our work shed light on the ability of ISIS to systematically and programmatically corroborate its agenda with remarkable coordinated activity on social media.

## Acknowledgements

The authors are grateful to Max Abrahms (Northeastern University) for useful discussions, and to Alessandro Flammini and Onur Varol (Indiana University) for their support in collecting the Twitter dataset. This work has been partly funded by the Office of Naval Research (ONR), grant no. N15A-020-0053. This research is also based upon work supported in part by the Office of the Director of National Intelligence (ODNI), Intelligence Advanced Research Projects Activity (IARPA). The views and conclusions contained herein are those of the authors and should not be interpreted as necessarily representing the official policies, either expressed or implied, of ODNI, IARPA, ONR, or the U.S. Government. The U.S. Government had no role in study design, data collection and analysis, decision to publish, or preparation of the manuscript. The U.S. Government is authorized to reproduce and distribute reprints for governmental purposes notwithstanding any copyright annotation therein.

## Tables, Figures & Lists

**Table I**

| Violence | | | Theological | | |
|---|---|---|---|---|---|
| *Stem* | *Translation* | *Frequency* | *Stem* | *Translation* | *Frequency* |
| قتل | To kill | 88732 | خلف | Caliph | 80664 |
| جهد | Jihad | 66268 | حسب | Judgment Day | 76116 |
| شهد | Martyr | 38027 | حمد | To Thank | 70131 |
| عرك | Fight | 36195 | دين | Religion | 67908 |
| حذف | Delete | 34718 | كبر | To call for | 56710 |
| حرب | War | 29030 | وحد | To unite and it is usually used to express belief in one God | 44225 |
| قصف | Bomb | 27534 | شيخ | Shiekh | 37630 |
| فجر | "Fajr" or dawn but also mean to explode | 24149 | رحم | To have Mercy | 37369 |
| فتح | Stem for military conquest in a religious sense | 22927 | رسل | Messenger | 32847 |
| | | | ولي | To rule or to be appointed to rule | 30352 |
| | | | شرع | Stem for Sharia | 25916 |
| | | | سور | Verse or wall | 22664 |

**Table II**

| | Names | | | Sectarian | |
|---|---|---|---|---|---|
| *Stem* | *Translation* | *Frequency* | *Stem* | *Translation* | *Frequency* |
| دولة_الإسلامية | Islamic State (ISIS in Arabic/how they liked to be called) | 111892 | رفض | To refuse, although here it would be a derogatory term for Shias | 66449 |
| اخبار_الخلافة | The Caliphate News | 65340 | صفي | Stem for a derogatory term for Shias | 40939 |
| ولة_الخلافة | The Caliphate State | 52864 | كفر | Disbelief or Infidel | 32967 |
| دعش | Stem for ISIS | 30336 | رتد | Can be the stem for apostasy, also used as derogatory term for adherents of others sects and religions | 24884 |
| اسد | Literally means lion, but here probably referring to Assad, whose name means lion in Arabic | | | | |
| غرب | West | 26475 | | | |
| عمر | Omar (name) or age | 23504 | | | |
| عرب | Arabs | 22328 | | | |
| عاصفة_الحزم | Decisive Storm- military operation led by Saudi Arabic in Yemen | 22153 | | | |

**Table III**v

| Category | Date | Tweet |
|---|---|---|
| *Names* | 1/20/2014 | الشعب العربي انقسم إلى قسمين: قسم اكل الصدأ رأسه وهو يصدق هرطقات الاعلام وعلماء التلفاز. والقسم الاخر بدأ يبحث بنفسه ليتبين له الحق |
| | | **Translation**: The Arab Nation is divided into two sections: one, its head is eaten by rust, this section believes in the lies of the media and so-called "scholars" of the television and the other section, started looking for the truth by itself |
| | 5/5/2015 | الوقائع على الارض توكد ان بشار الاسد يتراجع على كل الجبهات وهزيمته أصبحت واضحة جدا, بعدها ينتقل الصراع #اللاذقية ومن... |
| | | **Translation**: # Latakia and form the facts on the ground attest that Bashar al-Assad is pulling back on all fronts and his defeat has become very clear, after that, the struggle |
| *Sectarian* | 4/12/2015 | هذا هم المسلمين يجاهدون الشيعة بالعراق وحكامك يدعمون الشيعة عليهم وش الفرق بين الحوثي الرافضي والعراقي الرافضي؟ |
| | | **Translation**: These are the Muslims who fight (Jihad in the verb tense) Iraq's Shias and your leaders support the Shias against them (presumingly, the Sunnis), so what the difference (in Iraqi dialect) between a Rafidi Houthivi and a Rafidi Iraqi. |
| | 5/29/2015 | سنضرب أي حسينية من لايعرف أن المناطق الشرقية في السعودية هي عقر دار روافض جزيرة العرب لايحق له أن يكون محلل إعلامي |
| | | **Translation**: We will attack any Hussainiya,vii who does not know that the eastern parts of Saudi Arabia is the stronghold of the Arabian Peninsula's (Saudi Arabia) Rafidah does not deserve to be a media analyst |
| *Theological* | 8/11/2014 | وليعلم العالم ليس معنى أنه دخل الكعبة فهو ليس بزنديق، قد كسر الصنم، هذه صفعة لكل من يعتقد إن خائن الحرمين ولي أمر ولا يجوز الخروج عليه |
| | | **Translation**: Let the world know, the fact that he entered the Kaaba does not change the fact that he is a Zindiqviii, the idol has been broken, this is a slap on the face for anybody that believes that rebelling against the "traitor of the two holy places" (a word play on the title of the Saudi Kings, "Custodian of the Two Holy Mosques") and the Legal guardian (ruler in Islamic Fiqh) cannot be permitted (according to Islamic law) |
| | 4/24/2015 | قال عليه الصلاة والسلام: من لزم الاستغفار جعل الله له من كل هم فرجاً، ومن كل ضيق مخرجاً، ورزقه من حيث لا يحتسب |
| | | **Translation**: (This is a saying by Prophet Mohamed) Peace and prayers be upon him said: "If anyone constantly seeks pardon (from Allah), Allah will appoint for him a way out of every distress and a relief from every anxiety, and will provide sustenance for him from where he expects not." (Hadith) |
| *Violence* | 1/1/2014 | من الضروري جدا فتح كافة الجبهات في كافة المحافظات المنتفضة و ذلك للتخفيف على الرمادي واقضيتها الثائرة ولتشتيت قوات نوري العميل |

|  | **Translation**: It is very necessary to open all fronts in all revolting/rebellious provinces and that is in order to lessen the burden on Ramadi (city in central Iraq) and its rebellious neighborhoods and to distract/divide the forces of Nouri the traitor (referring to Nouri el-Maliki, the former prime minister of Iraq). |
|---|---|
| 6/6/2015 | نزف لكم نبأ استشهاد الأخ المهاجر أبو أنس روقة .. جندي من جنود الدولة الإسلامية وذلك في معارك شرق ولاية الرقة |
|  | **Translation**: We announce (with pleasure) the news of brother Abu Anas Warqa' (the migrant) martyrdom, a solider among the soldiers of the Islamic State, and his martyrdom occurred in the battles of east Al-Raqqah Province. |

**Table IV**

| Categories | Number | Out of the Total (%) | Out of the Categorized (%) |
|---|---|---|---|
| **Names** | 168663 | 8.7 | 15.5 |
| **Other** | 257670 | 13.3 | 23.7 |
| **Sectarian** | 74731 | 3.9 | 6.9 |
| **Theological** | 387090 | 20 | 35.6 |
| **Violence** | 197950 | 10.2 | 18.2 |
| **None** | 846912 | 43.8 | |
| **Total number of Tweets = 1,933,016; Total number of Tweets Categorized = 1,086,104** | | | |

**Table V**

| Names + Categories | Number | Out of the Total Names Dataset (%) |
|---|---|---|
| **Theological** | 50098 | 34.8 |
| **Violence** | 40829 | 28.38 |
| **Other** | 33664 | 23.4 |
| **Sectarian** | 19490 | 13.5 |
| **Total number of Names + Categories Tweets = 144,081** | | |

**Table VI**

| Name | Date | Event Description |
|---|---|---|
|  | 1/5/2014 | Fallujah taken by ISIS |
|  | 1/14/2014 | Raqqa becoming the capital of ISIS |
|  | 4/1/2015 | Iraqi government takes over Takrit |
| **Crucifixions** | 5/1/2014 | Crucifixions in Raqqa |
| **Mosul and Takrit Captured** | 6/10/2014 | Mosul and Takrit taken by ISIS |
| **Iran Deploys** | 6/12/2014 | Iran deploys forces to fight ISIS in Iraq, and helps Iraqi troops regain control of most of Tikrit. |
| **Iraq (USA Support)** | 6/18/2014 | Iraq asks the United States to conduct airstrikes against ISIS |
| **Caliphate** | 6/28/2014 | ISIS announces the establishment of a caliphate and rebrands itself as the "Islamic State." |
| **Shaer Gas Battle** | 7/17/2014 | ISIS storms the Shaer gas field and kills 270 people. |
| **Sinjar Captured** | 8/2/2014 | ISIS conquers Kurdish towns of Sinjar and Zumar, forcing thousands of Yazidi civilians to flee their homes. |
| | 8/7/2014 | President Obama announces the beginning of air strikes against ISIS in Iraq to defend Yazidi citizens stranded in Sinjar |
| | 9/2/2014 | Leaders from ISIS and its jihadist rival, Jabhat al Nusra, meet in Atareb to discuss joining forces. No formal merger or cooperation between the groups is established, but ISIS reportedly sent fighters to help the Nusra Front's assault on Harakat Hazm, a Western-backed moderate rebel group |
| **Kobani Caputred** | 9/9/2014 | ISIS advances on the Syrian border town of Kobani and thousands of refugees flee into Turkey. |
| | 9/22/2014 | ISIS spokesman Abu Muhammad al Adnani calls for attacks on citizens of the United States, France and other countries involved in the coalition to destroy the group |
| | 9/23/2014 | The United States launches its first air strikes against ISIS in Syria |
| | 10/7/2014 | The United States significantly ramps up airstrikes in and around Kobani to counter ISIS advances. |
| | 1/7/2015 | Two gunmen, Saïd and Chérif Kouachi, attack the offices of French satirical newspaper Charlie Hebdo in Paris, killing 11 people. A third assailant, Amedy Coulibaly, carried out a synchronized attack on a kosher supermarket, taking hostages and killing four people. Coulibaly reportedly declared allegiance to the Islamic State. |
| | 1/26/2015 | Kurdish fighters, with the help of U.S. and coalition airstrikes, force out ISIS militants from the Syrian border town of Kobani after a four-month battle |
| **Violent Acts** | 2/15/2015 | -ISIS releases a video of Jordanian military pilot Moaz al Kasasbeh being burned alive. (2/4/2015)<br>-Libyan militants allied to ISIS release a video showing the beheading of 21 Egyptian Christians, who had been kidnapped on January 12. Egypt launches airstrikes in Libya in retaliation. (2/15/2015)<br>-ISIS militants abduct at least 200 Assyrian Christians in northeastern Syria. The U.S.-led coalition launches airstrikes in the same area. (2/25/2015) |
| **Takrit Liberated** | 3/2/2015 | The Iraqi government launched a massive military operation to recapture Tikrit with 30,000 Iraqi soldiers and backed by air force. |
| **Yazidis Released** | 4/8/2015 | ISIS releases more than 200 captive Yazidis, most of whom had been held captive in northwestern Iraq since mid-2014. |
| **Christian Killings** | 4/19/2015 | ISIS posts a video showing militants from its Libyan branch executing dozens of Ethiopian Christians. |
| **Palmyra Captured** | 5/20/2015 | ISIS seizes the ancient Syrian city of Palmyra. |
| | 5/21/2015 | ISIS militants take full control of Sirte, Libya-Muammar Qaddafi's hometown. |
| **Shite Attack** | 5/22/2015 | ISIS claims responsibility for the suicide attacks on a Shiite mosque in eastern Saudi Arabia, which killed 21 people and injured more than 100. |
| | 6/26/2015 | ISIS fighters kill at least 145 civilians in an attack on Kobani, Syria. The same day, ISIS-linked militants attacked a Shiite mosque in Kuwait, killing 27 people and injuring more than 200. |

**Table VII**

| *Stem* | *Translation* |
|---|---|
| في | In |
| من | From |
| أن | That |
| علي | On |
| إلي | To |
| التي | Which |
| عن | About |
| لا | No |
| ما | What |
| هذا | This (Male) |
| هذه | This (Female) |
| كان | It was |
| مع | With |
| و | And |
| ذلك | That |
| بين | Between |
| لم | Did not |
| بعد | After |
| كل | All |
| الذي | Which |

**Figure I**

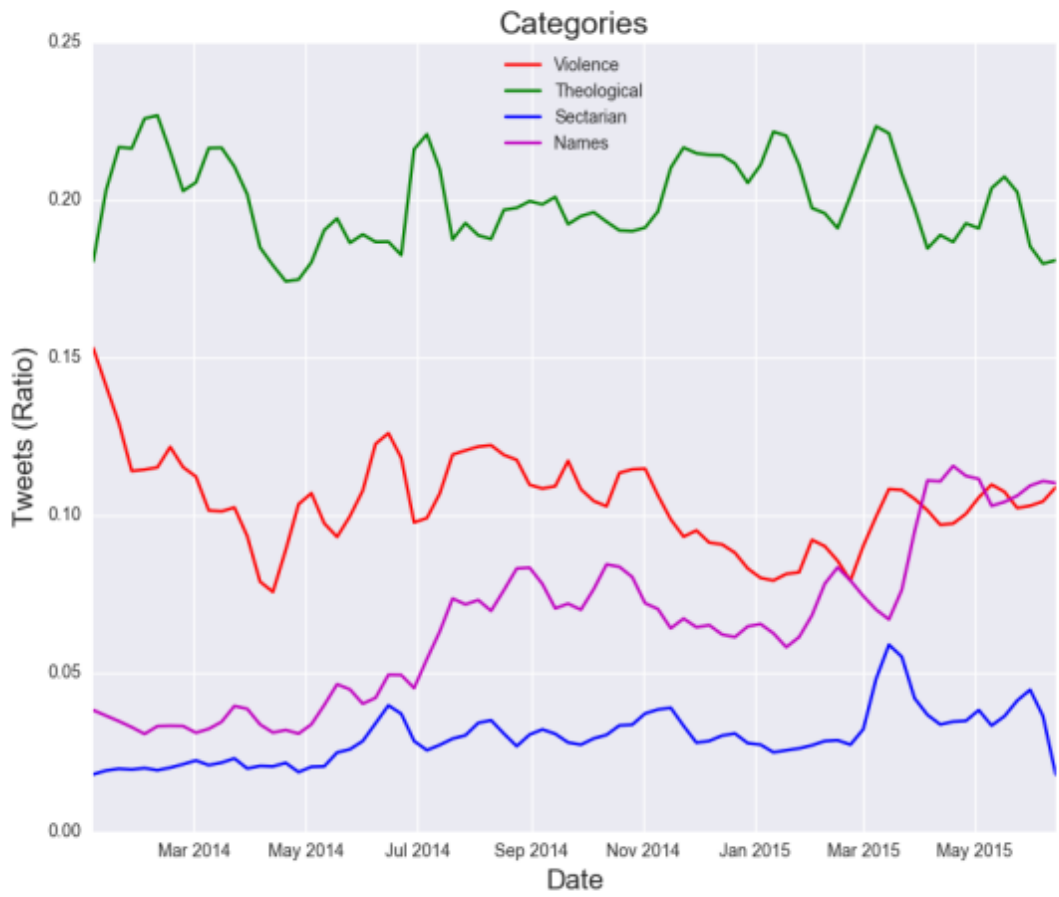

**Figure II**

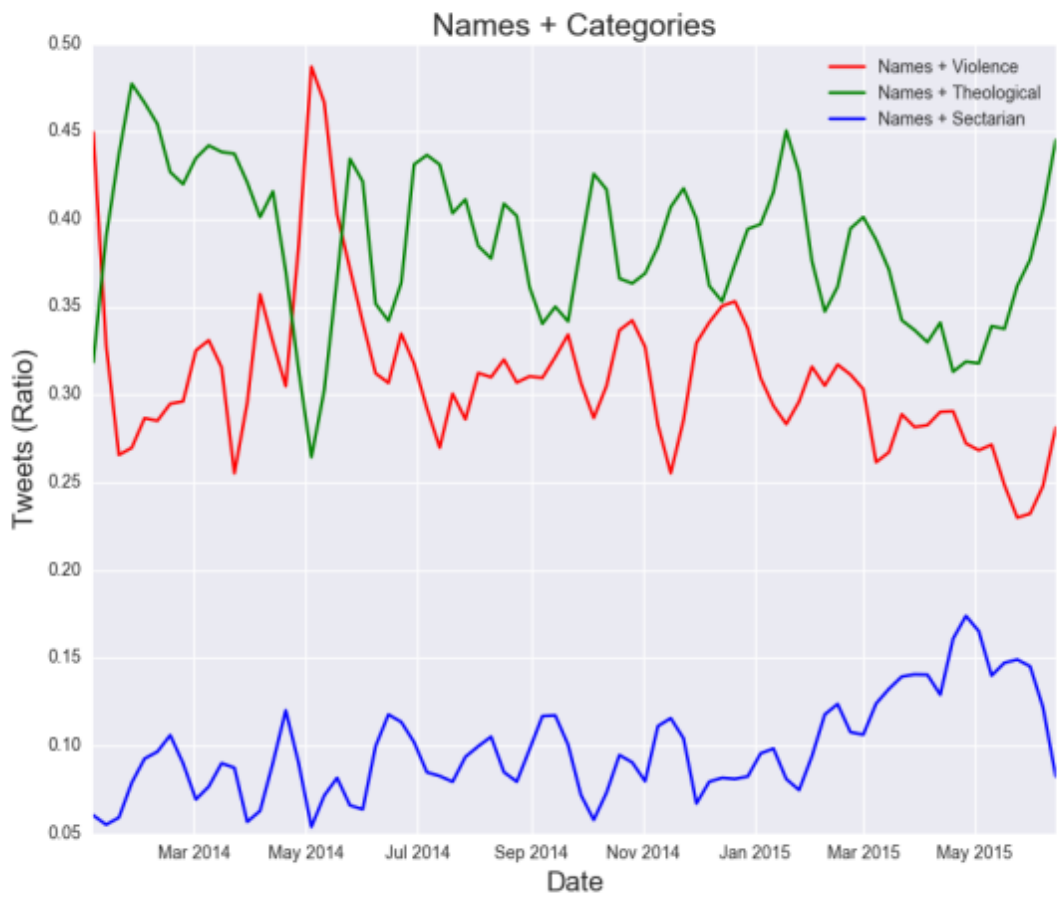

**Figure III**

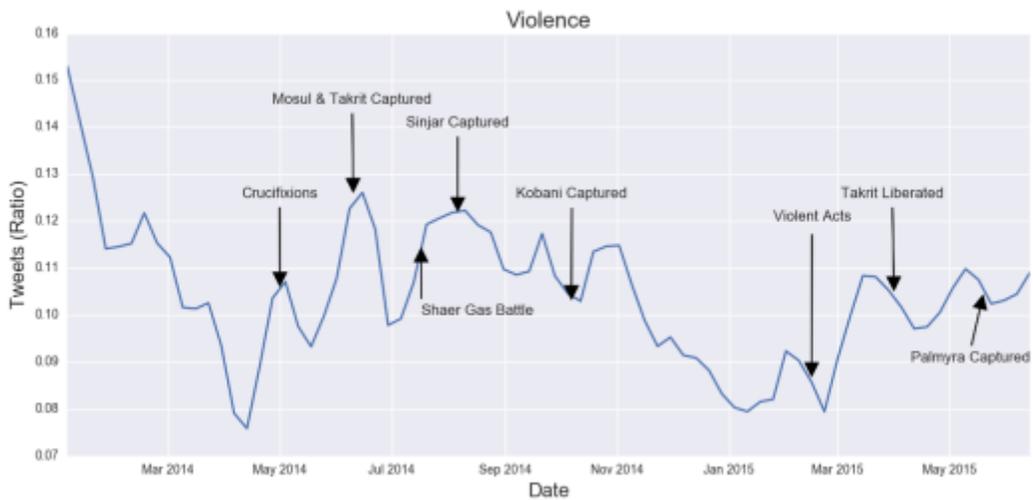

**Figure IV**

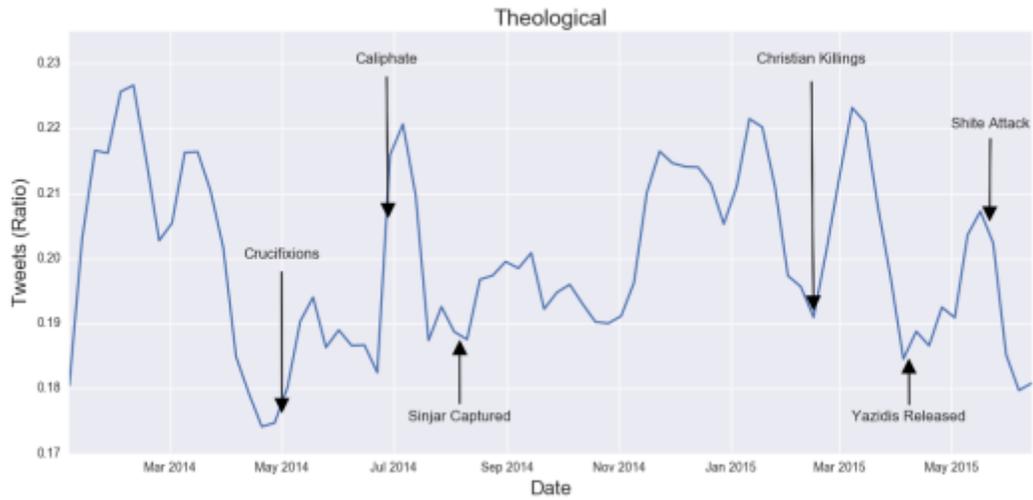

**Figure V**

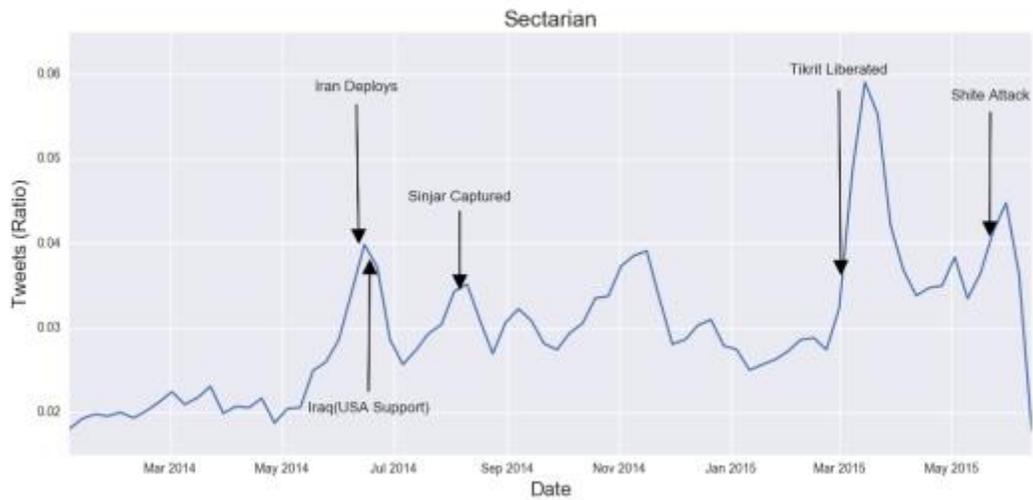

**Figure VI**

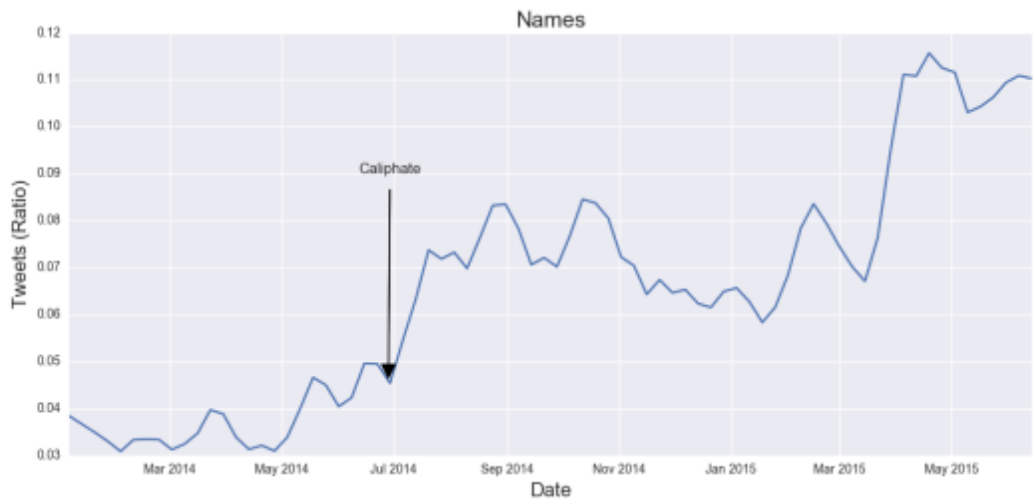

**Figure VII**

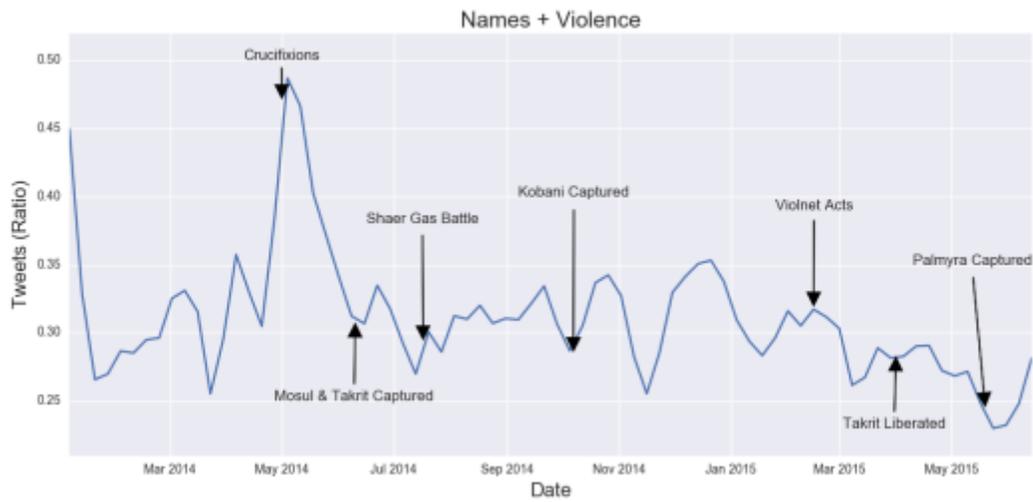

**Figure VIII**

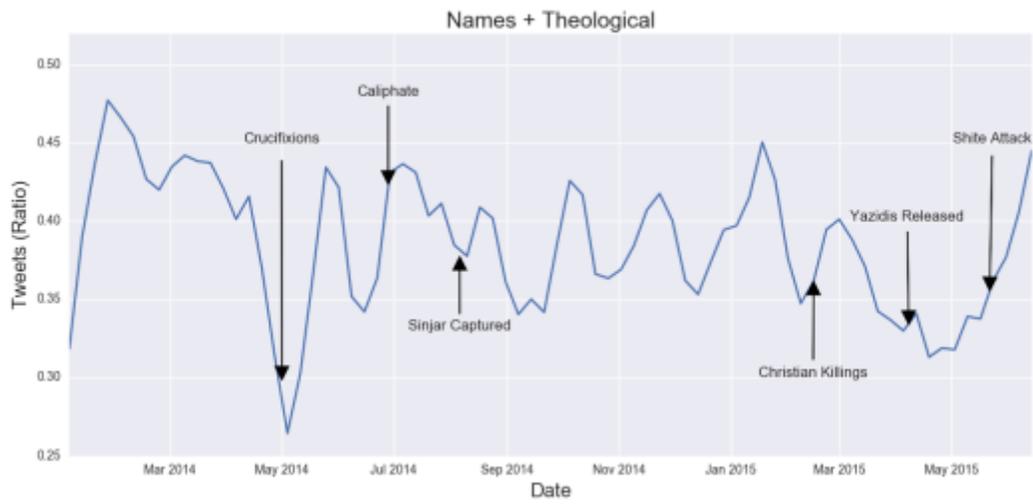

**Figure IX**

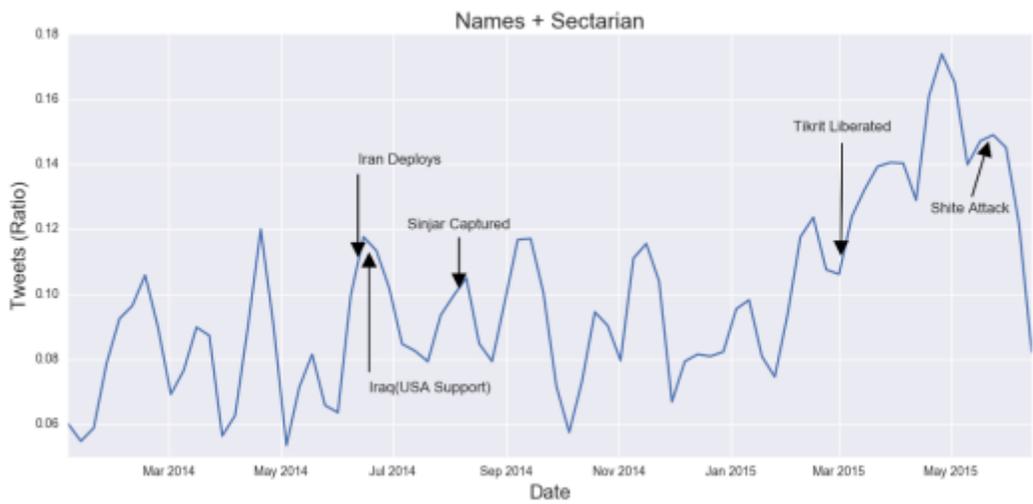

# Notes

---

[i] The bag-of-words model is a simplifying representation used in natural language processing where a piece of text is represented as the bag (multiset) of its words, disregarding grammar and even word order but keeping multiplicity.

[ii] "Five" in the code

[iii] For more on the subject, read chapter eight "The AQ-ISIS War" in (Stern & Berger, 2015)

[iv] For more on the subject, read Milestones (Qutb 1964)

[v] We used (Glen 2016) as a guide for choosing what events to put in this table.

[vi] The Houthis are a Zaidi Shia-led religious-political movement based in Northern Yemen and is in engaged in ongoing battles with the current Yemeni government and Saudi Arabia.

[vii] A gathering congregation hall for Shia commemoration ceremonies to commemorate the martyrdom of Imam Hussein.

[viii] Term used in the medieval ages for Muslims who "strayed from the right path and in believing in monotheism", could be punishable by death.  (Lewis 1993)